\documentstyle[preprint,aps,eqsecnum]{revtex}
\def\mpla{{Mod.\ Phys.\ Lett.\ }{\bf A}}
\def\npb{{Nucl.\ Phys.\ }{\bf B}}

\def\plb{{Phys.\ Lett.\ }{\bf B}}
\def\prd{{Phys.\ Rev.\ }{\bf D}}
\def\prl{Phys.\ Rev.\ Lett.\ }

\begin{document}

\preprint{\vbox{\hbox{Liverpool Preprint LTH470}   
\hbox{hep-th/9912105}
\hbox{Revised Version}}}

\title{\Large \bf  $T$-Duality and Conformal Invariance at Two Loops}

\author{S.~Parsons}

\address{Dept. of Mathematical Sciences,
University of Liverpool, Liverpool L69 3BX, UK}

\maketitle

\begin{abstract}  
{We show that the conformal invariance conditions for a general $\sigma$-model
with torsion are invariant under $T$-duality through two loops.}
\end{abstract}

\setcounter{footnote}{0}
\renewcommand{\thefootnote}{\arabic{footnote}}
\setcounter{section}{0}
\section{Introduction}
Duality invariance is a tremendously powerful concept in string theory. 
One of the earliest forms of duality to be recognised was that now known as 
$T$-duality\cite{busch}. This 
acts to transform the background fields of a $\sigma$-model
so as to map one conformally-invariant background (or string vacuum)
into another conformally-invariant background, at least at lowest order. 
In fact
the $\sigma$-model and its dual should be equivalent, again at least at lowest 
order\cite{rocek}\cite{balog1}.
The duality
can be understood as a consequence of an isometry of the theory; upon gauging
the isometry, by performing the path integral over the gauge field and the 
path integral over a lagrange multiplier in different orders, one obtains 
two equivalent descriptions of the theory with backgrounds related by the
duality. This duality is straightforwardly checked at lowest order in $\alpha'$;
conformal invariance requires the vanishing of ``$B$''-functions, one for each 
background field (the metric $g$, the antisymmetric tensor $b$ and
the dilaton $\phi$), which are  
related to the $\beta$-functions for the corresponding background fields in a 
way which we will describe later. One can check that the duality  
transformations on backgrounds for which the $B$-functions vanish lead to 
dual backgrounds which also have vanishing $B$-functions. In fact, this property
is equivalent to requiring that the $B$-functions are form-invariant under 
duality, and hence satisfy certain consistency
conditions which were derived in Ref.~\cite{haag}\cite{balog2}
(see also Ref.~\cite{schi}).
However, it is far less clear whether, and if so how, the $T$-duality is
maintained at higher orders in $\alpha'$. In Ref.~\cite{tseyt} for a
restricted background, and in Ref.~\cite{kal} for the general case, it
has been found that the two-loop string effective action can be made duality 
invariant by a redefinition of the background fields (see also 
Ref.~\cite{meiss}). Now the $B$-functions 
are related to the string effective action, and so this is grounds
for hoping that the $B$-functions will also be invariant. 
However, this is not a {\it fait accompli}, as was explained in Refs.\cite{haag}
\cite{haag1}. 
Certainly the invariance 
of the one-loop effective action guarantees the invariance of the one-loop 
$B$-functions. On the other hand, the relation between 
the action and the $B$-functions is more complicated at 
higher orders. Moreover, once one has decided that the fields need to be
redefined at higher orders to maintain duality invariance, then the 
$B$-functions will also be modified in a non-trivial way, because field 
redefinitions lead to changes in the $\beta$-functions. For these reasons it 
seems desirable to investigate explicitly
whether the field redefinitions which make the 
action duality-invariant also lead to duality-invariant $B$-functions. This
verification was carried out in Ref.~\cite{haag} for the restricted case of 
Ref.~\cite{tseyt}. 
Here we shall carry out the analysis for a different, complementary case which 
we believe displays most of the features of the full general situation. 

\section{Duality at First Order}

The two-dimensional non-linear $\sigma$-model is defined by the action 

\begin{equation}
 S = \Huge \frac{1}{4\pi\alpha'} \int \large  d^{\:2} \sigma \left[ \sqrt{
\gamma} g_{\mu \nu}( X)\partial_{A} X^{\mu}
\partial_{B} X^{\nu} \gamma^{ A B} + ib_{\mu \nu}( X)
\partial_{A} X^{\mu} \partial_{B} X^{\nu} \epsilon^{
A B} +\sqrt{\gamma}\ R^{(2)} \phi( X ) \right],
\end{equation}where $g_{\mu\nu}$ is the metric, $b_{\mu\nu}$ is the antisymmetric 
tensor field often referred to as torsion, and $\phi$ is the dilaton. 
The indices  $\mu,\nu$ run over 1,\ldots$D$$+$$1$. ${\gamma}_{AB}$ 
is the metric on the two-dimensional world sheet, with $A$,$B$ = 1,2. 
$\gamma$ = det ${\gamma}_{AB}$ , $\epsilon_{AB}$ is the two-dimensional
alternating 
symbol and $R^{(2)}$ the worldsheet Ricci scalar. Note that $b_{\mu\nu}$ is 
only defined up to a gauge transformation
$b_{\mu\nu} \mapsto b_{\mu\nu} + {\nabla_{[\mu}}{\zeta}_{\nu]}$.
Conformal invariance 
of the $\sigma$-model requires the vanishing of the Weyl anomaly 
coefficients $B^{g}$  , $B^{b}$   and $B^{\phi}$ (we will refer to these
as the $B$-functions), which are defined 
as follows\cite{ark}: 
\begin{equation}
B^{g}_{\mu \nu} = {\beta}^{g}_{\mu \nu} + 2 {\alpha}' {\nabla}
_{\mu} {\nabla}_{\nu} \phi + {\nabla}_{ ( \mu} {S}_{\nu )} ,
\end{equation} 
\begin{equation}
B^{b}_{\mu \nu} = {\beta}^{b}_{\mu \nu} + {\alpha}' H^{\rho}
_{\; \, \mu \nu} {\nabla}_{\rho} \phi + \frac{1}{2}H^{\rho}_{\; \, 
\mu \nu}S_{\rho},
\end{equation}
\begin{equation}
B^{\phi} = \beta^{\phi} +{\alpha}' {\nabla}^{\rho}
\phi {\nabla}_{\rho} \phi + \frac{1}{2}  {\nabla}^{\rho} \phi S_{\rho}.
\end{equation}
Here $\beta^{g}$  , $\beta^{b}$   and $\beta^{\phi}$ are the renormalisation 
group $\beta$-functions for the $\sigma$-model, and $H_{\mu\nu\rho}$ is
the field strength tensor for $b_{\mu\nu}$, defined by $H_{\mu\nu\rho}$
= $3{\nabla}_{[\mu} b_{\nu\rho]}$ = ${\nabla}_{\mu} b_{\nu\rho} 
+ \hbox{cyclic}$. 
The vector $S^{\mu}$ arises in the process of defining the trace 
of the energy-momentum tensor as a finite composite operator, 
and can be computed perturbatively. It will be sufficient to
assume $S^{\mu} = (S^{0},0)$.
At one loop we have 
\begin{equation}
{\beta}^{g(1)}_{\mu \nu} = R_{\mu \nu} - \frac{1}{4} H_{\mu \rho \sigma}
H_{\nu}^{\;\; \rho \sigma},
\end{equation}
\begin{equation}
{\beta}^{b(1)}_{\mu \nu} = - \frac{1}{2} {\nabla}_{\rho} H^{\rho}_{\;\;
\mu \nu},
\end{equation}
\begin{equation}
{\beta}^{\phi(1)} =- \frac{1}{2} \Box \phi - \frac{1}{24} H^{2},
\end{equation}
where $H^{2}$ = $H_{\mu \nu \rho}H^{\mu \nu \rho}$. (Here and henceforth we set
$\alpha'=1$.)

The conformal invariance conditions are equivalent to the
equations of motion of a string effective action
\begin{equation}
\Gamma = \Huge\int \large d^{D+1} X \sqrt{g} L({\lambda}_{M})
\end{equation}
where $ {\lambda}_{M} \equiv (g_{\mu\nu},b_{\mu\nu},\phi) $.
More explicitly, we have to leading order
\begin{equation}
{\Gamma}^{(1)} = \Huge\int \large d^{D+1} X \sqrt{g} e^{-2 \phi}
\left[ R(g) + 4 ( \nabla \phi)^{2} - \frac{1}{12} H_{\mu \nu \rho}
H^{\mu \nu \rho} \right].
\end{equation}
From this action it can be shown that
\begin{equation}
\frac{\partial {\Gamma}^{(1)}}{\partial \lambda_{M}} = 2\sqrt{g} e^{-2 \phi}
K^{(0)}_{MN} B^{(1)}_{N}
\end{equation}
where
\begin{equation}
K^{(0)}_{MN} \equiv \left( \begin{array}{ccc} g_{\mu}^{\;\rho} g_{\nu}^{\;
\sigma}& 0 & -  g_{\mu \nu} \\ 0 & - g_{\mu}^{\;\rho} g_{\nu}^{\;\sigma}
& 0 \\ 0 & 0 & 8 \\  \end{array} \right)
\end{equation}
and we have a basis for $B_{M}$ such that $B_{M} =
(B^{g}_{\mu\nu},B^{b}_{\mu\nu}, {\tilde{B}}^{\phi})$ and 

\begin{equation}
{\tilde{B}}^{\phi} = B^{\phi} - \frac{1}{4} g^{\mu\nu} B^{g}_{\mu\nu}.
\label{eq:1.12}
\end{equation}

We now consider the dual $\sigma$-model. 
This involves introducing an abelian isometry in the target space
background of the model such that one can perform duality
transformations. Background fields will now be locally independent of
the coordinate $\theta \equiv X^{0}$ and locally dependent on $X^{i}\;,i
= 1,\dots,D$. The $\sigma$-model action now reads as  
\begin{eqnarray}
 \large S  & = & \Huge\int \large d^{\:2} \sigma \frac{1}{4 
\pi \alpha'} \left[\sqrt{\gamma} \gamma^{ A B}  ( g_{00}( X^{k}) \partial_{A}
\theta \partial_{B} \theta  + 2 g_{0i}(X^{k})\partial_{A} \theta 
\partial_{B}X^{i} \nonumber  + g_{ij}(X^{k}) 
\partial_{A} X^{i} \partial_{B}X^{j}) \right. \nonumber \\
 & &\left.+ \sqrt{\gamma} R^{(2)} \phi( X^{k}) ) 
+i \epsilon^{A B} ( 2b_{oi}(X^{k})\partial_{A} \theta
\partial_{B}X^{i} +b_{ij}(X^{k}) \partial_{A} X^{i}
\partial_{B}X^{j} ) \right]
\end{eqnarray}
The classical $T$-duality transformations act on the background fields
$\{g_{\mu \nu},b_{\mu \nu}\}$ to give dual background fields
$\{{\tilde{g}}_{\mu\nu},{\tilde{b}}_{\mu \nu}\}$ given by\cite{busch}:
\[{\tilde{g}_{00}} = \frac{1}{\Huge g_{00}},\]
\[{\tilde{g}_{0i}} = \frac{b_{0i}}{{g}_{00}}, \: \;
{\tilde{b}_{0i}}
= \frac{g_{0i}}{{g}_{00}},\]
\begin{equation}
          {\tilde{g}_{ij}} = g_{ij} - \frac{g_{0i}g_{0j}-b_{0i}b_{0j}}
{g_{00}},
\end{equation}
          \[{\tilde{b}_{ij}} = b_{ij} - \frac{g_{0i}b_{0j}-
b_{0i}g_{0j}}{g_{00}}.\]
The origin of the the two distinct models lies in the order in which one
performs simple parts of the path integral.  
We shall deal with the transformation on the dilaton later.

We now parametrize the metric and torsion tensors in terms of reduced 
fields which appear in the Kaluza-Klein reduction to 
$D$-dimensions\cite{haag2}\cite{haag}\cite{kal}.
\begin{equation}
g_{\mu \nu} 
 =\left( \begin{array}{cc} g_{00} & g_{0j} \\ g_{i0} & g_{ij} \\
\end{array} \right) =
\left( \begin{array}{cc} a & a v_{j} \\ a v_{i} &
{\bar{g}}_{ij} +a v_{i} v_{j} \\ \end{array} \right),
\end{equation}

\begin{equation}
b_{\mu \nu} 
= \left( \begin{array}{cc} b_{00} & b_{0j} \\ b_{i0} & b_{ij} \\
\end{array} \right)  
= \left( \begin{array}{cc} 0 & w_{j}\\ -w_{i} & b_{ij}\\
\end{array} \right).
\end{equation}
This choice simplifies the  form of the classical transformations to 
\begin{eqnarray}
a \mapsto  \frac{1}{a} ,\;\;\;\ v_{i} \leftrightarrow w_{i}, \nonumber
\\ b_{ij} \mapsto {\tilde{b}}_{ij} = b_{ij} + w_{i}v_{j} -w_{j}v_{i}.
\label{eq:T}
\end{eqnarray}

Note that ${\bar{g}}_{ij}$ is unchanged under duality. It has been 
shown that a further simplification in the context of conformal 
invariance conditions can be employed, since the transformation
properties of the one loop $B$-functions are manifest when mapped to tangent
space\cite{haag}. That is, 
we construct a vielbein, $e_{a}^{\;\;\mu}$, such that 

\begin{equation}
\delta_{ab} = e^{\;\;\mu}_{a} e^{\;\;\nu}_{b} g_{\mu\nu}.
\end{equation}
In fact we can choose a block diagonal form
\begin{equation}
e^{\;\;\mu}_{a} = \left( \begin{array}{cc} e^{\;\;0}_{\hat{0}} 
& e^{\;\;i}_{\hat{0}}\\ e_{\alpha}^{\;\;0}  & e_{\alpha}^{\;\;i}\\
\end{array} \right)
= \left( \begin{array}{cc} \frac{1}{\sqrt{a}} & 0 \\
- v_{\alpha} & {\bar{e}}_{\alpha}^{\;\;i}\\
\end{array} \right).  
\end{equation}
with $v_{\alpha} \equiv {\bar{e}}_{\alpha}^{\;\;i}v_{i}$.
We can now define the tangent space anomaly coefficients as:
\begin{equation}
B^{g}_{ab} = e^{\;\;\mu}_{a} e^{\;\;\nu}_{b} B
^{g}_{\mu\nu}\;\; ,\;\; B^{b}_{ab} = e^{\;\;\mu}_{a} 
e^{\;\;\nu}_{b} B^{b}_{\mu\nu},
\end{equation}
\begin{equation}
B^{\tilde{g}}_{ab} = e^{\;\;\mu}_{a} e^{\;\;\nu}_{b}
B^{\tilde{g}}_{\mu\nu}\;\; ,\;\; B^{\tilde{b}}
_{ab} = e^{\;\;\mu}_{a}e^{\;\;\nu}_{b}
B^{\tilde{b}}_{\mu\nu}.
\end{equation}
These coefficients transform as
 \[ B^{\tilde{g}}_{\hat{0} \hat{0}} = - 
B^{g}_{\hat{0} \hat{0}}, \]
\begin{equation}
B^{\tilde{g}}_{\hat{0} \alpha} = B^{b}_
{\hat{0} \alpha} \; \; , \; \; B^{\tilde{b}}_{\hat{0}
\alpha} = B ^{g}_{\hat{0} \alpha},
\end{equation}  
\[ B^{\tilde{g}}_{\alpha \beta} = B^{g}_
{\alpha \beta} \; \; , \; \; B^{\tilde{b}}_{\alpha
\beta} = B^{b}_{\alpha \beta}. \]
With the exception of $B^{b}_{\alpha \beta}$, we find that we
can express each tangent space $B$-function in terms of 
a $B$-function for one reduced field,
up to factors of $g_{00}$. 
\begin{eqnarray}
B^{g}_{\hat{0}\hat{0}} &=& e_{\hat{0}}^{\;\;0}e_{\hat{0}}^{\;\;0}
B^{g}_{00} =\frac{1}{a} B^{a} \nonumber \\
&=& \frac{1}{a} {\beta}^{a} + a^{k}{\bar{\nabla}}_{k} \phi,  
\end{eqnarray}
\begin{eqnarray}
B^{g}_{\hat{0}\alpha} &=&  e_{\hat{0}}^{\;\;0} {\bar{e}}^
{\;\;i}_{\alpha}a B^{v}_{i} = {\bar{e}}^{\;\;i}_{\alpha} \sqrt{a}
B^{v}_{i} \nonumber \\
&=& {\bar{e}}^{\;\;i}_{\alpha} \sqrt{a} \left[ \beta^{v}_{i} -
F_{i}^{\;\;k} \bar{\nabla}_{k} \phi + \frac{1}{2} \bar{\nabla}_{i} S^{0},
\right]
\end{eqnarray}
\begin{eqnarray}
B^{b}_{\hat{0}\alpha} &=& e_{\hat{0}}^{\;\;0}{\bar{e}}^{\;\;i}_{\alpha}
B^{w}_{i} = {\bar{e}}^{\;\;i}_{\alpha} \frac{1}{\sqrt{a}} B^{w}_{i}
\nonumber \\
&=&  {\bar{e}}^{\;\;i}_{\alpha} \frac{1}{\sqrt{a}} \left[ \beta^{w}_{i}
- G_{i}^{\;\;k} \bar{\nabla}_{k} \phi \right],
\end{eqnarray}
\begin{eqnarray}
B^{g}_{\alpha \beta} &=& {\bar{e}}^{\;\;i}_{\alpha}     
{\bar{e}}^{\;\;j}_{\beta} B^{\bar{g}}_{ij} \nonumber \\
&=& {\bar{e}}^{\;\;i}_{\alpha} {\bar{e}}^{\;\;j}_{\beta} \left[ 
\beta^{\bar{g}}_{ij} +2 \bar{\nabla}_{i} \bar{\nabla}_{j} \phi \right],
\end{eqnarray}
\begin{eqnarray}
B^{b}_{\alpha \beta} &=& {\bar{e}}^{\;\;i}_{\alpha}     
{\bar{e}}^{\;\;j}_{\beta} B^{b^{\star} }_{ij} \nonumber \\
&=& {\bar{e}}^{\;\;i}_{\alpha}
{\bar{e}}^{\;\;j}_{\beta} \left[ \beta^{b^{\star} }_{ij} + \hat{H}^{k}
_{\;\;ij} \bar{\nabla}_{k} \phi -\frac{1}{2} G_{ij} S^{0} \right]. 
\end{eqnarray}
We define 
\begin{equation}
B^{b^{\star} }_{ij} = B^{b}_{ij} -{v}_{i}B^{w}_{j} + {v}_{j}B^{w}_{i}.
\end{equation}

The one loop $B$-functions for the reduced
tensors  are 
\begin{equation}
B^{a(1)} = -\frac{a}{2} {\bar{\nabla}}
_{i}a^{i} + a a^{i}{\bar{\nabla}}_{i}\Phi + \frac{1}{4} (a^{2}F_{km}
F^{km}-G_{km}G^{km}),
\end{equation}
\begin{equation}
B^{v(1)}_{i} = 
 - F_{i}^{\; \; k}( {\bar{\nabla}}_{k} \Phi-\frac{1}{2} 
a_{k}) + \frac{1}{2} {\bar{\nabla}}^{k} F_{ik}+ \frac{1}{4a}
 {\hat{H}}_{ikm}G^{km} + \frac{1}{2} {\bar{\nabla}}_{i} S^{0},
\end{equation} 
\begin{equation}
B^{w(1)}_{i} =
 - G_{i}^{\; \; k}( {\bar{\nabla}}_{k} \Phi +
\frac{1}{2}a_{k}) + \frac{1}{2} {\bar{\nabla}}^{k} G_{ik}
+ \frac{a}{4} {\hat{H}}_{ikm} F^{km},
\end{equation}
\begin{equation}
B^{\bar{g}(1)}_{ij} = 
 {\bar{R}}_{ij} - \frac{1}{4}a_{i}a_{j} - 
\frac{1}{2}(a F_{ik}F_{j}^{\;\;k} + \frac{1}{a} G_{ik}G_{j}^{\;\;k})-
 \frac{1}{4} {\hat{H}}_{ikm} {\hat{H}}_{j}^{\;\; km} + 2 {\bar{\nabla
}}_{i}{\bar{\nabla}}_{j} \Phi, 
\end{equation}
\begin{equation}
B^{b^{\star}(1)}_{ij} = 
 -  \frac{1}{2} {\bar{\nabla}}^{k} {\hat{H}}
_{kij} + {\hat{H}}_{kij} {\bar{\nabla}}^{k} \Phi - \frac{1}{2}G_{ij}S^{0},
\end{equation}
where $a_{i}= \partial_{i} \ln a$, $F_{ij} = 2 {\bar{\nabla}}_{[i}
v_{j]}$, $G_{ij} = 2 {\bar{\nabla}}_{[i}w_{j]} = -H_{0ij}$, and
$\hat{H}_{ijk} = H_{ijk} +3v_{[i}G_{jk]}$. All barred tensors and
covariant derivatives only have dependence on $\bar{g}_{ij}$,  i.e. that
part of metric that is invariant under the classical $T$-duality 
transformations. Under the mapping (\ref{eq:T})
we have $a_{i} \mapsto -a_{i}$, $F_{ij} \leftrightarrow G_{ij}$
and ${\hat{H}}_{ijk}$ invariant. $\Phi$ is the reduced dilaton defined by  
\begin{equation}
\Phi = \phi - \frac{1}{4} \ln a.
\end{equation}
$\Phi$ is another invariant under the duality transformation and this is
seen to be case if one combines the shift in $\phi$ needed to keep the
one loop action invariant\cite{kal} with the transformation on $a$. The
corresponding $B$-function can be calculated, though
it need not be worked out in full detail. (\ref{eq:1.12}) 
simplifies to
\begin{equation}
{\tilde{B}}^{\phi}= B^{\Phi} -\frac{1}{4} B^{g}_{\alpha\alpha}.
\end{equation}  
This is manifestly invariant and indeed this  fact is closely related
to the invariance of the reduced one loop string effective action which
can be similarly expressed\cite{kal}
\begin{eqnarray}
\Gamma_{R}^{(1)} = \Huge\int\large d^{d} X \sqrt{\bar{g}} e^{-2 \Phi}
\left[ \bar{R}(\bar{g}) + 4 {\bar{\nabla}}^{i} \Phi {\bar{\nabla}}_{i}
\Phi - \frac{1}{4}a^{i}a_{i} \right. \nonumber \\
- \left. \frac{1}{4} a F_{ij} F^{ij}-\frac{1}{4a}G_{ij}G^{ij} -
\frac{1}{12} {\hat{H}}_{ijk}{\hat{H}}^{ijk} \right].
\end{eqnarray}
From a reduced effective action one would expect the equations of
motion for the reduced tensors to reproduce the appropriate conformal
invariance conditions. In fact, we have the following (see (A4)---(A7)):
\begin{equation}
e_{ \hat{0}}^{\;\;\mu}e_{ \hat{0}}^{\;\;\nu}  
\left. \frac{\partial \Gamma_{R}}{\partial g^{\mu \nu}} \right|_{b_{\mu\nu}} 
= - 2a \frac{\partial \Gamma_{R}}{\partial a},  
\end{equation}  
\begin{equation}
e_{\alpha}^{\;\; \mu} e_{\beta}^{\;\; \nu} \left. \frac{\partial \Gamma_{R}}
{ \partial g^{\mu \nu}} \right|_{b_{\mu\nu}} = {\bar{e}}_{\alpha}^
{\;\; i} {\bar{e}}_{\beta}^{\;\; j}\frac{\partial \Gamma_{R}}{ \partial 
{\bar{g}}^{ij}},
\end{equation}
\begin{equation}
e_{ \hat{0}}^{\;\;\mu}e_{\alpha}^{\;\; \nu}  \left. \frac{\partial \Gamma_{R}}
{\partial g^{\mu \nu}} \right|_{b_{\mu\nu}} = - {\bar{e}}_
{\alpha}^{\;\; i} {e}_{ \hat{0}} ^{\;\;0} \frac{\partial \Gamma_{R}}
{ \partial v^{i}},
\end{equation}
\begin{equation}
e_{ \hat{0}}^{\;\;\mu}e_{\alpha}^{\;\; \nu} \left. \frac{\partial \Gamma_{R}}
{\partial b^{\mu \nu}} \right|_{g_{\mu\nu}} = {\bar{e}}_
{\alpha}^{\;\; i} {e}_{\hat{0}}^{\;\;0} a \frac{\partial \Gamma_{R}}
{ \partial w^{i}},
\end{equation}
\begin{equation}e_{\alpha}^{\;\; \mu} e_{\beta}^{\;\; \nu} \left.
\frac{\partial \Gamma_{R}}{ \partial b^{\mu \nu}} \right|_{g_{\mu\nu}} = 
{\bar{e}}_{\alpha}^{\;\; i} {\bar{e}}_{\beta}^{\;\; j}\frac{\partial
\Gamma_{R}}{ \partial b^{ij}}.
\end{equation}
The validity of this set of equations at leading order is clear 
upon inspection of (2.29)---(2.33) and (2.36), together with
(2.2)---(2.7) and (2.9).From now on we drop the 
$R$ subscript, since all quantities will be assumed to be reduced.
 
\section{Duality at Second Order}

We shall now proceed to illustrate the transformation properties of the
next to leading order conformal invariance conditions. We would like to show 
that the two-loop conformal invariance conditions behave under duality as 
in (2.22), but we shall see that we have to redefine the fields in order to 
achieve this. There are 
two ways in which we might proceed. The more direct 
approach is to calculate the $B$-functions explicitly for the fields we are
considering. These expressions are very complicated, so we abstain from
this. The indirect method is to consider the corresponding result to (2.10)
at second order
\begin{equation}
\frac{\partial {\Gamma}^{(2)}}{\partial {\lambda}_{M}} = 2\sqrt{g} e^{-2\phi}
\left[ K^{(0)}_{MN} B_{N}^{(2)} + K_{MN}^{(1)} B^{(1)}_{N} \right],
\end{equation} 
where ${\Gamma}^{(2)}$ is given in Ref.~\cite{jj}.
$K_{MN}^{(1)}$ can be read off the following expressions\cite{jj},
\begin{equation}
\frac{1}{\sqrt{g} e^{-2 \phi}} \frac{ \partial \Gamma^{(2)}}
{\partial \phi}= 8 {\tilde{B}}^{\phi(2)} + 2 B^{b(1)}_{\rho
\sigma}B^{b(1)\rho \sigma} + 2B^{g(1)}_
{\rho\sigma}B^{g(1)\rho \sigma} - 2
({\tilde{B}}^{\phi(1)} )^{2}, 
\end{equation}
\begin{eqnarray}
\frac12\frac{1}{\sqrt{g} e^{-2 \phi}} \frac{ \partial \Gamma^{(2)}}
{\partial g^{\mu \nu}}
 &=&  B^{g(2)}_{\mu \nu} - B^{g(1)}_{\mu \rho}
B^{g(1) \rho}_{\nu} + R_{\mu \;\;\;\; \nu}^{\;\;
\rho \sigma} B^{g(1)}_{ \rho \sigma}
+  {\tilde{B}}^{\phi(1)}B^{g(1)}_{\mu \nu} \nonumber \\
&& - B^{b(1)}_{\mu \rho}B^{b(1) \rho}_{\nu}
+ \frac{3}{4}H_{\mu \rho \tau} \nonumber  H^{\tau}_{\;\; \nu \sigma}   
B^{g(1) \rho \sigma} +  \frac{1}{2}{\nabla}_{( \mu} H_{\nu)}^{\;\;
\rho \sigma} B^{b(1)}_{\rho \sigma} \nonumber \\
&& + g_{\mu \nu} \left[ - 4{\tilde{B}}^{\phi(2)} -
B^{b(1)}_{\rho \sigma}B^{b(1)\rho \sigma}
-  B^{g(1)}_{\rho\sigma}B^{g(1)
\rho \sigma} +  ( {\tilde{B}}^{\phi(1)} )^{2} \right],   
\end{eqnarray}
\begin{eqnarray}
\frac12\frac{1}{\sqrt{g} e^{-2 \phi}} \frac{ \partial \Gamma^{(2)}}
{\partial b^{\mu \nu}}
&=& - B^{b(2)}_{\mu \nu} + R_{\rho \mu \nu \sigma} B^{b(1) \rho \sigma}
- \frac{1}{4} H_{\tau \mu \nu} H^{\tau \rho
\sigma}B^{b(1)}_{\rho \sigma}
+ \frac{1}{4} H_{\mu \tau \rho} H_{\nu \;\; \sigma}^{\;\; \tau}
B^{b(1)\rho \sigma} \nonumber \\
 &&- 2 B^{g(1)}_{[\mu | \rho |}{B}^{b(1) \rho}_
{\nu]} -\frac{1}{2} {\nabla}^{\sigma} H^{\rho}_{\;\; \mu \nu}
B^{g(1)}_{ \rho \sigma} - {\tilde{B}}
^{\phi(1)} B^{b(1)}_{\mu \nu}.
\end{eqnarray}   
With these relations, we could show the properties under duality of
$B_{M}^{(2)}$ if we knew the properties of the second order action 
and those of $K_{MN}^{(1)}$. After all we are well informed to the 
behaviour of $B^{(1)}$. However we know already that $\Gamma^{(2)}$ is
not the appropriate action. In fact  in Ref.~\cite{kal} it was proved that a 
shift in the reduced fields is required to obtain an invariant second 
order action. The shift in the one loop reduced effective action is then
\begin{eqnarray}
\delta {\Gamma}^{(1)} = \Huge\int\large  d^{d} X \sqrt{\bar{g}} e^{-2
\Phi}
\left[ - \frac{1}{2} a^{i} {\nabla}_{i} \delta a  
 - \frac{a}{2}\left(\frac{1}{2a} \delta a F^{ij} F_{ij}
+ F^{ij} \delta F_{ij} \right) \right. \nonumber \\ + \left. \frac{1}{2a}
\left( \frac{1}{2a} \delta a  G^{ij} G_{ij} - G^{ij} \delta G_{ij}\right) -
\frac{1}{6} {\hat{H}}^{ijk}\delta {\hat{H}}_{ijk} \right].
\end{eqnarray}
The authors of Ref.~\cite{kal} find 
the required leading order corrections to the reduced fields to be
\begin{equation}
 \delta a =  a a_{i} a^{i}
+ \frac{1}{8} a^{2} F^{ij} F_{ij} + \frac{1}{8} G^{ij} G_{ij},  
\end{equation}
\begin{equation}
\delta v_{i} = - \frac{1}{4} F_{i}^{\;\;k} a_{k} -
\frac{1}{8a} {{\hat{H}}}_{ikm} G^{km},
\end{equation}
\begin{equation}
\delta w_{i} = - \frac{1}{4} G_{i}^{\;\;k} a_{k} + \frac{1}{8} a 
{{\hat{H}}}_{ikm} F^{km},
\end{equation}
\begin{equation} 
\delta {\hat{H}}_{ijk}= - \frac{3}{2} {\bar{\nabla}}_{[i}\left( G_{j}^{\;\;m}
F_{k]m} \right) + 3 F_{[ij} \delta w_{k]} + 3 G_{[ij} \delta v_{k]}.
\end{equation}
We now have at our disposal an invariant action which we shall write
as 
\begin{equation}
{\Gamma'}^{(2)} = \Gamma^{(2)} + \delta \Gamma^{(1)}.
\end{equation}
However the consequence of (3.6)---(3.9) in (3.1) is not just that we have to
replace $\Gamma^{(2)}$ with $\Gamma'^{(2)}$  but also that the
two-loop $\beta$-functions change, leading consequently to a modified 
$K$-matrix. The new $\beta$-functions are in general given by:
\begin{equation}
\beta'_{M} = \beta_{M} - \delta \beta_{M}
\end{equation}
where  
\begin{equation}
\delta \beta_{M}(\lambda) = \delta \lambda . \frac{\delta}{
 \delta \lambda} \beta_{M} (\lambda) - \mu \frac{d}{d \mu} \delta
{\lambda}_{M}.
\end{equation}
The precise details of how we apply (3.12) will be given in the Appendix, and
moreover we will leave for later the definition of the redefined $K$-matrix.
We now
seem to have all the required apparatus to achieve our task. In the
expressions that follow, after all functional derivatives have been
taken such as in an equation of motion or $\delta {\beta}_{M}$, 
we will 
consider the simplified theory with $a=1$ and $\Phi=0$. (The
complementary case with general $a$, but with $\Phi=b_{ij}=0$, was considered in
Ref.~\cite{haag}.)  
 
We first turn our attention to the dilaton. This turns out to be the
simplest calculation, mainly because there is no shift in the reduced 
dilaton, $\Phi$, and consequently the equations of motion are unaltered.
By comparing (2.9) and (2.36) we see that the functional dependence
on $\phi$ in the unreduced action is the same as on $\Phi$ in the
reduced action. Therefore (3.2) becomes 
\begin{eqnarray}
\frac{1}{\sqrt{g} e^{-2 \Phi}} \frac{ \partial \Gamma'^{(2)}}
{\partial \Phi}  &=& 8 {\tilde{B}}^{\Phi(2)} + 2 B^{b(1)}_{ab}B^{b(1)}_{ab}
+ 2B^{g(1)}_{ab}B^{g(1)}_{ab} - 2
({\tilde{B}}^{\Phi(1)} )^{2} \nonumber \\
&=& 8 {\tilde{B}}^{\Phi(2)} + 4 B^{b(1)}_{\hat{0}\alpha} 
B^{b(1)}_{\hat{0}\alpha} + 2 B^{b(1)}_{\alpha\beta} 
B^{b(1)}_{\alpha\beta} + 4 B^{g(1)}_{\hat{0}\alpha}B^{g(1)}_{\hat{0}\alpha}
\nonumber \\&+& 2 B^{g(1)}_{\hat{0}\hat{0}} B^{g(1)}_{\hat{0}\hat{0}}
+ 2 B^{g(1)}_{\alpha\beta}B^{g(1)}_{\alpha\beta} -
2({\tilde{B}}^{\Phi(1)} )^{2},
\end{eqnarray} 
where 
\begin{equation}
{\tilde{B}}^{\Phi}= {B}^{\Phi} - \frac{1}{4} {\bar{g}}^{ij} B^{\bar{g}}_{ij} 
= {\tilde{B}}^{\phi}
\end{equation}
and ${\tilde{B}}^{\phi}$ is given in (2.35). Given that the
$B^{(1)}$-functions satisfy (2.22) and that 
$\tilde{B}^{\Phi(1)}$ is invariant, we
conclude that that ${\tilde{B}}^{\Phi(2)}$ is invariant under 
duality.

We now turn our attention to $B^{g}_{\alpha\beta}$. The fundamental
equation for us to deal with is
\begin{equation}
\frac12e_{\alpha}^{\;\;\mu} e_{\beta}^{\;\;\nu} \frac{\partial
\Gamma^{(2)}}{\partial g^{\mu\nu}} = e_{\alpha}^{\;\;\mu}
e_{\beta}^{\;\;\nu} B^{g(2)}_{\mu\nu} + e_{\alpha}^{\;\;\mu}
e_{\beta}^{\;\;\nu} K^{g}_{\mu\nu}.
\end{equation}

We can rewrite (3.15) using our analysis in the previous section of the
equations of motion and conformal invariance conditions of the reduced
fields, namely (2.26) and (2.38). This gives us
\begin{equation}
\frac12{\bar{e}}_{\alpha}^{\;\; i}{\bar{e}}_{\beta}^{\;\; j} \frac{\partial
\Gamma^{(2)}}{\partial \bar{g}^{ij}} =
{\bar{e}}_{\alpha}^{\;\;i}{\bar{e}}_{\beta}^{\;\; j}
B_{ij}^{\bar{g}(2)} + e_{\alpha}^{\;\;\mu} e_{\beta}^{\;\;\nu}
K^{g}_{\mu\nu}.
\end{equation}  
Correspondingly we can write 
\begin{equation}
\frac12{\bar{e}}_{\alpha}^{\;\; i}{\bar{e}}_{\beta}^{\;\; j} \frac{\partial
\delta \Gamma^{(1)}}{\partial \bar{g}^{ij}} =
{\bar{e}}_{\alpha}^{\;\;i}{\bar{e}}_{\beta}^{\;\; j}
\delta B_{ij}^{\bar{g}(2)} + e_{\alpha}^{\;\;\mu} e_{\beta}^{\;\;\nu}
X^{g}_{\mu\nu}.
\end{equation}
where $\delta B_{ij}^{\bar{g}(2)}$ is given in the appendix.
Finally we combine (3.16) and (3.17) to find 
\begin{equation}
{\bar{e}}_{\alpha}^{\;\; i}{\bar{e}}_{\beta}^{\;\; j} \left[
{B}_{ij}^{\bar{g}'(2)} \right]
 ={\bar{e}}_{\alpha}^{\;\; i}{\bar{e}}_{\beta}^{\;\; j}
 \frac{\partial}{ \partial
\bar{g}^{ij}}[ \Gamma'^{(2)}] + {K}^{g'}_{\alpha \beta}
\end{equation}
 
where
\begin{eqnarray}
{K}^{g'}_{\alpha \beta} &=& -e_{\alpha}^{\;\;\mu} e_{\beta}^{\;\;\nu}   
\left[ K^{g}_{\mu\nu}+ X^{g}_{\mu\nu} \right] \nonumber \\
&=&B^{g(1)}_{\alpha\gamma}B^{g(1)}_{\beta\gamma} + B^{g(1)}_{\alpha \hat{0}}
B^{g(1)}_{\beta\hat{0}} -{\tilde{B}}^{\Phi} B^{g(1)}_{\alpha \beta}   
+ B^{b(1)}_{\alpha \hat{0}}B^{b(1)}_{\beta \hat{0}} +B^{b(1)}_{\alpha\gamma}
B^{b(1)}_{\beta\gamma} \nonumber \\
&+& B^{g(1)}_{\hat{0} \hat{0}} {\bar{e}}_{\alpha}^{\;\;i}
{\bar{e}}_{\beta}^{\;\;j} \left[ \frac{1}{2} F_{ik} F_{j}^{\;\;k}
- \frac{1}{2} G_{ik} F_{j}^{\;\;k} \right] \nonumber \\
&+& B^{g(1)}_{\hat{0} \gamma} {\bar{e}}_{\alpha}^{\;\;i}   
{\bar{e}}_{\beta}^{\;\;j} {\bar{e}}_{\gamma}^{\;\;k} \left[ \bar{\nabla}
_{i}F_{kj} + \frac{1}{2} G_{ni}{\hat{H}}^{n}_{\;\;jk} + \frac{1}{2}
G_{nj}{\hat{H}}^{n}_{\;\;ik} \right] \nonumber \\
&+& B^{b(1)}_{\hat{0} \gamma}{\bar{e}}_{\alpha}^{\;\;i}{\bar{e}}_{\beta}
^{\;\;j} {\bar{e}}_{\gamma}^{\;\;k} \left[ \bar{\nabla}_{i}G_{kj} +
\frac{1}{2} F_{ni}{\hat{H}}^{n}_{\;\;jk} + \frac{1}{2}
F_{nj}{\hat{H}}^{n}_{\;\;ik} \right] \nonumber \\
&+&B^{g(1)}_{\gamma\delta}{\bar{e}}_{\alpha}^{\;\;i}{\bar{e}}_{\beta}^
{\;\;j} {\bar{e}}_{\gamma}^{\;\;k} {\bar{e}}_{\delta}^{\;\;m} \left[
{\bar{R}}_{ikjm} - \frac{3}{4} F_{ik}F_{jm} - \frac{3}{4} G_{ik}  
G_{jm} - \frac{3}{4} {\hat{H}}_{ikn}{\hat{H}}^{n}_{\;\;jm} \right]
\nonumber \\  
&+&B^{b(1)}_{\gamma\delta}{\bar{e}}_{\alpha}^{\;\;i}{\bar{e}}_{\beta}^
{\;\;j} {\bar{e}}_{\gamma}^{\;\;k} {\bar{e}}_{\delta}^{\;\;m} \left[
\frac{1}{4} {\bar{\nabla}}_{i} {\hat{H}}_{jkm} + \frac{1}{4}
{\bar{\nabla}}_{j} {\hat{H}}_{ikm} \right] \nonumber \\
&+&{\delta}_{\alpha \beta} \left[  {\tilde{B}}^{\Phi(2)} +
B^{b(1)}_{\hat{0} \gamma} B^{b(1)}_{\hat{0} \gamma} + B^{b(1)}_{\gamma
\delta}B^{b(1)}_{\gamma \delta} + B^{g(1)}_{\hat{0}\hat{0}} B^{g(1)}_
{\hat{0}\hat{0}}\right. \nonumber \\ 
&+& \left.B^{g(1)}_{\hat{0} \gamma} B^{g(1)}_{\hat{0} \gamma}
+ B^{g(1)}_{\gamma \delta} B^{g(1)}_{\gamma \delta} -({\tilde{B}}^
{\Phi(1)})^{2} \right] 
\end{eqnarray}   

So we have an invariant equation of motion and, from (2.22), an
invariant ${K}^{g'}_{\alpha \beta}$ and hence an invariant ${B}^
{g'(2)}_{\alpha\beta}$. 

We now follow a similar argument to prove the transformation properties
of $B^{g(2)}_{\hat{0}\hat{0}}$. Once again we combine the equation of
motion of $\Gamma^{(2)}$ ,
\begin{equation}
- a \frac{\partial {\Gamma}^{(2)}}{\partial a}  
 =  \frac{1}{a}
\left[ B^{a(2)} + K^{g}_{00} \right],
\end{equation}

with its correction
\begin{equation}
- a \frac{\partial \delta \Gamma^{(1)}}{\partial a} 
 =  \frac{1}{a} \left[ \delta B^{a(2)}+ X^{a} \right] 
\end{equation}

We find

\begin{equation}
\frac{1}{a} {B}^{a'(2)} = - a \frac{\partial {\Gamma'}^{(2)}}{\partial a}
  + \frac{1}{a} {K}^{a'} ,
\end{equation}
 
where
\begin{eqnarray}
{K}^{a'}& =&
- \left[ X^{a} + K^{g}_{00} \right]  \nonumber \\
&=& {B}^{g(1)}_{\alpha \beta} {\bar{e}}_{\alpha}^{\;\;i}
{\bar{e}}_{\beta}^{\;\;j} \left[ \frac{1}{2} F_{ik} F^{\;\;k}_{j} 
- \frac{1}{2} G_{ik} G^{\;\;k}_{j} \right] + {B}^{b(1)}_
{\alpha \beta} {\bar{e}}_{\alpha}^{\;\;i} {\bar{e}}_{\beta}^{\;\;j}
\left[ \frac{1}{4} F_{ik} G^{\;\;k}_{j} - \frac{1}{4} G_{ik}
F^{\;\;k}_{j}  \right]- B^{a(1)}{\tilde{B}}^{\Phi(1)}. \nonumber \\
\end{eqnarray}  
Since $a \mapsto \frac{1}{a}$ and $\frac{\partial \Gamma}{\partial a}
= -\frac{1}{a^2} \frac{\partial \Gamma}{\partial \left( \frac{1}{a}
\right)}$ we have
$-a \frac{\partial \Gamma}{\partial a} \mapsto a \frac{\partial
\Gamma}{\partial a}$. So both the equation of motion    
and $K$-matrix change sign under duality, and hence so does     
${B}_{\hat{0}\hat{0}}^{g'(2)}$.

The next check on duality concerns the mapping of $B^{g}_{\hat{0}
\alpha} \leftrightarrow B^{b}_{\hat{0}\alpha}$. For this case we will
need to compare two equations of motion and two $K$-matrices. For the
equations of motion, as can be seen in the appendix, we will deal with
functional derivatives with respect to $v_{i}$ and $w_{i}$
of $\Gamma$. In the case of the field equation for $v_{i}$ we keep 
${b}_{ij}$ and $w_{i}$ constant, while in the case of the field equation for 
$w_i$ we keep ${\tilde b}_{ij}$ and $v_i$ constant (where ${\tilde{b}}_{ij}$ 
was defined in (1.17)). Correspondingly, we
can write $\hat{H}_{ijk}$ either as 
\begin{equation}
\hat{H}_{ijk} = 3\bar{\nabla}_{[i}b_{jk]} + 3v_{[i}G_{jk]} 
\end{equation}
or as
\begin{equation}
\hat{H}_{ijk} = 3\bar{\nabla}_{[i} {\tilde{b}}_{jk]} + 3 w_{[i}F_{jk]}.
\end{equation}
(Note that this displays the duality invariance of $\hat{H}_{ijk}$.)
We shall also use these two forms of $\hat{H}_{ijk}$ when 
calculating $\delta B^{v}_{i}$ and $\delta B^{w}_{i}$ respectively. 
See the appendix for more details.

So for $v_{i}$ we have
\begin{equation}
 {e}_{ \hat{0}}^{\;\;0} {\bar{e}}_{\alpha}^{\;\; i}
\left[- \frac{1}{2} \frac{\partial\Gamma^{(2)}}{\partial v^{i}}
\right] =  {e}_{ \hat{0}}^{\;\;0}  {\bar{e}}_{\alpha}^{\;\; i}\left[
B^{v(2)}_{i} \right] + {e}_{\alpha}^ {\;\;\mu}
{e}_{\hat{0}}^{\;\;\nu} K^{g}_{\mu\nu},
\end{equation}
and a correction
\begin{equation}
 {e}_{ \hat{0}}^{\;\;0} {\bar{e}}_{\alpha}^{\;\; i}
\left[- \frac{1}{2} \frac{\partial \delta \Gamma^{(1)}}{\partial v^{i}} 
\right] =  {e}_{ \hat{0}}^{\;\;0}  {\bar{e}}_{\alpha}^{\;\; i}\left[  
\delta B^{v(2)}_{i} \right] + {e}_{\alpha}^ {\;\;\mu}  
{e}_{\hat{0}}^{\;\;\nu} X^{v}_{\mu\nu},
\end{equation} 
where
\begin{equation}
\delta B^{v(2)}_{i} = \delta \beta^{v(2)}_{i} +\frac{1}{2}
\bar{\nabla}_{i} \delta S^{0},
\end{equation}
with
\begin{equation}
\delta S^{0}= -\frac{1}{2}B^{b^{\star}}_{km}G^{km}. 
\end{equation}
We sum these to give
\begin{equation}
{e}_{ \hat{0}}^{\;\;0} {\bar{e}}_{\alpha}^{\;\; i} \left[
{B}^{v'(2)}_{i} \right]
=- \frac{1}{2} {e}_{ \hat{0}}^{\;\;0}{\bar{e}}_{\alpha}^{\;\; i} \left[
\frac{\partial}{\partial v^{i}} [ {\Gamma'}^{(2)}] \right]
+ {K}^{v'}_{ \hat{0} \alpha},
\end{equation}
where
\begin{eqnarray}
{K}^{v'}_{ \hat{0} \alpha} &=& - {e}_{\alpha}^ {\;\;\mu}  {e}_{\hat{0}}^
{\;\;\nu} \left[  X^{v}_{\mu\nu} + K^{g}_{\mu \nu} \right] \nonumber \\ 
&=&  B^{g(1)}_{\alpha \gamma}B^{g(1)}_{ \hat{0} \gamma} + B^{g(1)}_{\gamma   
\delta} {\bar{e}}_{\alpha}^{\;\;i} {\bar{e}}_{\gamma}^{\;\;k}
{\bar{e}}_{\delta}^{\;\;m} \left[ \frac{1}{2} {\bar{\nabla}}_{k}
F_{im} + \frac{1}{2} G_{kn} {\hat{H}}^{n}_{\;\;im} \right] \nonumber \\
&+&  B^{b(1)}_{\gamma \delta} {\bar{e}}_{\alpha}^{\;\;i}
{\bar{e}}_{\gamma}^{\;\;k}{\bar{e}}_{\delta}^{\;\;m} \left[
 \frac{1}{4} F_{in} {\hat{H}}^{n}_{\;\;km} + \frac{1}{4}
F_{kn} {\hat{H}}^{n}_{\;\;mi} \right] \nonumber \\
&+&  B^{g(1)}_{\hat{0} \gamma}{\bar{e}}_{\alpha}^{\;\;i} {\bar{e}}_
{\gamma}^{\;\;k} \left[\frac{1}{4} G_{in}G_{k}^{\;\;n} - \frac{1}{4}
F_{in}F^{\;\;n}_{k} \right] \nonumber \\
&+& B^{b(1)}_{\hat{0} \gamma}{\bar{e}}_{\alpha}^{\;\;i} {\bar{e}}_
{\gamma}^{\;\;k} \left[\frac{1}{4} G_{in}F_{k}^{\;\;n} - \frac{1}{4}
F_{in}G_{k}^{\;\;n} \right] -\frac{1}{8} {\bar{e}}_{\alpha}^{\;\;i} 
{\bar{\nabla}}_{i} \left(B^{b^{\star}}_{km}G^{km} \right). 
\end{eqnarray} 
For $w_{i}$ we have
\begin{equation}
 {e}_{ \hat{0}}^{\;\;0} {\bar{e}}_{\alpha}^{\;\; i} \left[
\frac{1}{2}\frac{\partial\Gamma^{(2)}}{\partial w^{i}} \right]
=-  {e}_{\hat{0}}^{\;\;0}  {\bar{e}}_{\alpha}^{\;\; i}\left[
B^{w(2)}_{i} \right] +{e}_{\alpha}^ {\;\;\mu} {e}_{\hat{0}}
^{\;\;\nu}K^{b}_{\mu\nu},                    
\end{equation}
and a correction
\begin{equation}
 {e}_{ \hat{0}}^{\;\;0} {\bar{e}}_{\alpha}^{\;\; i} \left[
\frac{1}{2}\frac{\partial \delta \Gamma^{(1)}}{\partial w^{i}} \right] 
=-  {e}_{\hat{0}}^{\;\;0}  {\bar{e}}_{\alpha}^{\;\; i}\left[\delta 
B^{w(2)}_{i} \right] +{e}_{\alpha}^ {\;\;\mu} {e}_{\hat{0}}
^{\;\;\nu}X^{w}_{\mu\nu}.
\end{equation}
We sum these to give
\begin{equation}
{e}_{ \hat{0}}^{\;\;0} {\bar{e}}_{\alpha}^{\;\; i} \left[
{B}^{w'(2)}_{i} \right]
=-\frac{1}{2} {e}_{ \hat{0}}^{\;\;0}{\bar{e}}_{\alpha}^{\;\; i}
\left[\frac{\partial}{\partial w^{i}} [ \Gamma'^{(2)}] \right]+
{K}^{w'}_{ \hat{0} \alpha},
\end{equation}
where 
\begin{eqnarray}
{K}^{w'}_{ \hat{0} \alpha} &=& {e}_{\alpha}^ {\;\;\mu} {e}_
{\hat{0}}^{\;\;\nu} \left[X^{w}_{\mu\nu}+ K^{b}_{\mu \nu} \right]
\nonumber \\
&=& B^{g(1)}_{\alpha \gamma} B^{b(1)}_{\hat{0} \gamma} + B^{g(1)}_
{\gamma \delta}
{\bar{e}}_{\alpha}^{\;\;i} {\bar{e}}_{\gamma}^{\;\;k} 
{\bar{e}}_{\delta}^{\;\;m} \left[ \frac{1}{2} {\bar{\nabla}}_{k} G_{im}
+ \frac{1}{2} F_{kn} {\hat{H}}^{n}_{\;\;im} \right] \nonumber \\
&+& B^{b(1)}_{\gamma \delta} {\bar{e}}_{\alpha}^{\;\;i} {\bar{e}}_
{\gamma}^{\;\;k}{\bar{e}}_{\delta}^{\;\;m} \left[ \frac{1}{4}
G_{in} {\hat{H}}^{n}_{\;\;km} + \frac{1}{4} G_{kn} {\hat{H}}^{n}_{\;\;
mi} \right] \nonumber \\
&+& B^{b(1)}_{\hat{0} \gamma} {\bar{e}}_{\alpha}^{\;\;i}{\bar{e}}_
{\gamma}^{\;\;k} \left[ \frac{1}{4} F_{in} F_{k}^{\;\;n}
- \frac{1}{4} G_{in} G_{k}^{\;\;n} \right] \nonumber \\
&+& B^{g(1)}_{\hat{0} \gamma} {\bar{e}}_{\alpha}^{\;\;i}{\bar{e}}_
{\gamma}^{\;\;k} \left[ \frac{1}{4} F_{in} G_{k}^{\;\;n} - \frac{1}
{4} G_{in} F_{k}^{\;\;n} \right] -\frac{1}{8} {\bar{e}}_{\alpha}^
{\;\;i}{\bar{\nabla}}_{i} \left( B^{b^{\star}}_{km} F^{km} \right).
\end{eqnarray}
It is readily apparent , given the dual nature of the equations of
motion and ${K}^{w'}_{\hat{0}\alpha} \leftrightarrow {K}^{v'}
_{\hat{0}\alpha}$, that ${B}^{b'(2)}_{\hat{0}\alpha} \leftrightarrow 
{B}^{g'(2)}_{\hat{0}\alpha}$.

We now consider $B^{b(2)}_{\alpha\beta}$. The equation of motion of
$\Gamma^{(2)}$ gives
\begin{equation}
\frac12{\bar{e}}_{\alpha}^{\;\;i}{\bar{e}}_{\beta}^{\;\;j} \frac{\partial
\Gamma^{(2)}}{\partial b^{ij}} = - {\bar{e}}_{\alpha}^{\;\;i}{\bar{e}}
_{\beta}^{\;\;j} B^{b^{\star}(2)}_{ij} + {e}_{\alpha}^{\;\;\mu}{e}_{\beta}
^{\;\;\nu} K^{b}_{\mu\nu}.
\end{equation}
Following from the previous calculations, one might expect our next step
is to write an equation  containing $\delta B^{b^{\star}}_{ij}$. However
this is not well defined since as already stated in (2.28), there is no
$b^{^{\star}}_{ij}$ such that one can compute $\mu\frac{d}{d\mu}b^{\star
}_{ij}$ to obtain $\beta^{b^{\star}}_{ij}$. Hence we cannot write down a
$\delta{b^{\star}}_{ij}$. But we still have
\begin{equation}
\frac12{\bar{e}}_{\alpha}^{\;\;i}{\bar{e}}_{\beta}^{\;\;j} \frac{\partial
\delta \Gamma^{(1)}} {\partial b^{ij}} = - e_{\alpha}^{\;\;\mu} 
e_{\beta}^{\;\;\nu} \delta B^{b(2)}_{\mu\nu} + e_{\alpha}^{\;\;\mu}
e_{\beta}^{\;\;\nu} X^{b^{\star}}_{\mu\nu}
\end{equation}
The details concerning the computation of $e_{\alpha}^{\;\;\mu}e_{\beta}
^{\;\;\nu} X^{b^{\star}}_{\mu\nu}$ are given in the appendix. So we now sum
(3.36) and (3.37) to give
\begin{equation}
{B}^{b'(2)}_{\alpha\beta} = -{\bar{e}}_{\alpha}^{\;\;i}{\bar{e}}_{\beta}^
{\;\;j} \frac{\partial{\Gamma'}^{(2)}}{\partial b^{ij}} +
{K}^{b^{\star\prime}}_{\alpha\beta},
\end{equation}
where 
\begin{equation}
{B}^{b'(2)}_{\alpha\beta} = {\bar{e}}_{\alpha}^{\;\;i}{\bar{e}}
_{\beta}^{\;\;j} \left[ B^{b^{\star}(2)}_{ij} -\frac{1}{2} G_{ij} \delta
S^{0} + \Lambda_{ij} \right] + e_{\alpha}^{\;\;\mu}
e_{\beta}^{\;\;\nu} \delta B^{b(2)}_{\mu\nu} 
\end{equation}
and
\begin{eqnarray}
{K}^{b^{\star\prime}}_{\alpha\beta} &=& {e}_{\alpha}^{\;\;\mu}{e}_{\beta}^{\;\;\nu}
\left[X^{b^{\star}}_{\mu\nu} + K^{b}_{\mu\nu} \right] \nonumber \\
&=& B^{b}_{\alpha\gamma} B^{g}_{\beta\gamma} - B^{g}_{\alpha\gamma}
B^{b}_{\beta\gamma} +B^{g}_{\hat{0}\hat{0}}
{\bar{e}}_{\alpha}^{\;\;i}{\bar{e}}_{\beta}^{\;\;j} \left[ F_{in}G_{j}
^{\;\;n} - G_{in}F_{j}^{\;\;n} \right] \nonumber \\
&+&B^{g}_{\hat{0}\gamma}{\bar{e}}_{\alpha}^{\;\;i}{\bar{e}}_{\beta}^{\;\;j}
{\bar{e}}_{\gamma}^{\;\;k} \left[ -\frac{1}{4}F_{in}
{\hat{H}}^{n}_{\;\;kj} +\frac{1}{4} F_{jn}{\hat{H}}^{n}_{\;\;ki}
+ \frac{1}{2} F_{kn}{\hat{H}}^{n}_{\;\;ij} \right] \nonumber \\
&+&B^{b}_{\hat{0}\gamma}{\bar{e}}_{\alpha}^{\;\;i}{\bar{e}}_{\beta}^{\;\;j}
{\bar{e}}_{\gamma}^{\;\;k} \left[ -\frac{1}{4}G_{in}
{\hat{H}}^{n}_{\;\;kj} +\frac{1}{4} G_{jn}{\hat{H}}^{n}_{\;\;ki}
+ \frac{1}{2} G_{kn}{\hat{H}}^{n}_{\;\;ij} \right] \nonumber \\
&+&B^{b}_{\gamma\delta}{\bar{e}}_{\alpha}^{\;\;i}{\bar{e}}_{\beta}^{\;\;j}
{\bar{e}}_{\gamma}^{\;\;k}{\bar{e}}_{\delta}^{\;\;m} \left[
{\bar{R}}_{ikmj} +\frac{1}{4} {\hat{H}}_{ink}{\hat{H}}^{n}_{\;\;mj}
-\frac{1}{4} {\hat{H}}^{n}_{\;\;ij}{\hat{H}}_{nkm} \right. \nonumber \\
&&+\left.\frac{1}{4}G_{ik} G_{jm} +\frac{1}{4}F_{ik} F_{jm} +\frac{1}{8}
F_{ij}F_{km} +\frac{1}{8} G_{ij}G_{km} \right] \nonumber \\
&+&B^{g}_{\gamma\delta}{\bar{e}}_{\alpha}^{\;\;i}{\bar{e}}_{\beta}^{\;\;j}
{\bar{e}}_{\gamma}^{\;\;k}{\bar{e}}_{\delta}^{\;\;m} \left[
-\frac{1}{2} {\bar{\nabla}}_{k}{\hat{H}}_{mij} \right],
\end{eqnarray}
where $\delta{S}^{0}$ is as
given in (3.31). $\Lambda_{ij}$ corresponds to the gauge
freedom of the torsion tensor,
\begin{equation}
\Lambda_{ij} = \frac{1}{4} \bar{\nabla}_{i} \left[ G_{j}^{\;\;k}
{\beta}^{v(1)}_{k} - F_{j}^{\;\;k} {\beta}^{w(1)}_{k} \right]
- i\leftrightarrow j.
\end{equation}
Although $b_{ij} \mapsto \tilde{b}_{ij}$ , the equations of motion for
these two fields are the same. Given that the  $K$-matrix (3.40)
is also invariant we conclude that ${B}^{b'(2)}_{\alpha\beta}$ is 
invariant too.

\section{Conclusions}
We have shown that the conformal invariance conditions are invariant for a 
model with non-vanishing torsion, but where we have set the reduced dilaton
$\Phi$ to zero, and taken 
$g_{00}=1$. Since the fields require to be
redefined, the conformal invariance conditions refer to a redefined
renormalisation scheme. Alternatively, as was done in Ref.~\cite{kal}, one 
could leave the fields, and hence the renormalisation scheme, unchanged, but
instead modify the duality transformations. The required consistency conditions 
for the conformal invariance conditions would then no longer be the simple ones
we use, as given for instance in (2.22); 
but our results of course guarantee that
these new consistency conditions will be satisfied through two loops.
  
Our results are complementary to 
those of Ref.~\cite{haag}, where the torsion was zero but a non-vanishing 
dilaton and non-constant $a$ were used. We believe our results display the
main features of the general calculation and a non-vanishing
dilaton and non-constant $a$ could be incorporated into our results without
changing their basic form. As a consequence of the indirect method of 
calculation, we have not explicitly computed the various conformal invariance 
conditions; but if desired they could be obtained quite straightforwardly from
our final results using
the explicit expression for $\Gamma^{\prime(2)}$ given in Ref.~\cite{kal}--for 
instance, ${B}_{ij}^{\bar{g}'(2)}$ could be obtained from Eqs.~(3.18), 
(3.19). 

\section*{\sc Acknowledgements}
S.P. was supported by a PPARC Graduate Studentship. Many thanks go to I.
Jack for numerous helpful conversations and indefatigable patience and 
enthusiasm. 
\appendix
\setcounter{equation}{0}
\section{\sc Equations of motion}
First we give $b^{\mu\nu}$
\begin{equation}
b^{0i}=\frac{1}{a}w^{i} - v^{k} {\tilde{b}}_{k}^{\;\;i},\;\;\; 
b^{ij}={\tilde{b}}^{ij}.
\end{equation}

We have the following relations, with $\Gamma$ as  the reduced action

\begin{eqnarray}
\left.\frac{\partial \Gamma}{\partial {\bar{g}}^{ij}}
\right|_{b_{\mu\nu},v_{i},a,\Phi}
&=& \frac{\partial \Gamma}
{\partial {g}^{kl}} \frac{\partial {g}^{kl}}{\partial {\bar{g}}^{ij}}
+ \frac{\partial \Gamma}{\partial{g}^{k0}} \frac{\partial{g}^{k0}} 
{\partial {\bar{g}}^{ij}} + \frac{\partial \Gamma}{\partial{g}^{00}}
\frac{\partial{g}^{00}}{\partial {\bar{g}}^{ij}} \nonumber \\
&=& \frac{\partial \Gamma}{\partial {\bar{g}}^{ij}} - \frac{\partial
\Gamma}{\partial{g}^{i0}}
v_{j} - \frac{\partial \Gamma}{\partial{g}^{j0}} v_{i}
+v_{i}v_{j}\frac{\partial
\Gamma}{\partial{g}^{00}}
\end{eqnarray}
\begin{eqnarray}
\left.\frac{\partial \Gamma}{\partial v^{i}} \right|_ 
{b_{\mu\nu},{\bar{g}}_{ij},a,\Phi} &=& \frac{\partial
\Gamma}{\partial{g}^{kl}} \frac{\partial {g}^{kl}}{\partial {v}^
{i}} + \frac{\partial \Gamma}{\partial{g}^{k0}} \frac{\partial{g}
^{k0}}{\partial v^{i}}+ \frac{\partial \Gamma}{\partial{g}^{00}}  
\frac{\partial{g}^{00}}{\partial v^{i}} \nonumber \\
&=& -  \frac{\partial
\Gamma}{\partial{g}^{i0}} + v_{i} \frac{\partial \Gamma}{\partial{g}^   
{00}}
\end{eqnarray}
\begin{eqnarray}
\left.\frac{\partial \Gamma}{\partial
a}\right|_{b_{\mu\nu},v_{i},{\bar{g}}
_{ij},\Phi} 
&=& \frac{\partial\Gamma}{\partial
{g}^{kl}} \frac{\partial {g}^{kl}}{\partial a} + \frac{\partial
\Gamma}{\partial{g}^{k0}} \frac{\partial{g}^{k0}}{\partial a}
+ \frac{\partial \Gamma}{\partial{g}^{00}} \frac{\partial{g}^{00}}
{\partial a}  \nonumber \\
&=& - \frac{1}{{a}^{2}}\frac{\partial \Gamma}
{\partial{g}^{00}}
\end{eqnarray}
\begin{eqnarray}
\left. \frac{\partial \Gamma}{\partial w^{i}} \right|_{g_{\mu\nu},
{\tilde{b}}_{ij}, \Phi} &=& \frac{\partial\Gamma}{\partial b^{0k}} 
\frac{\partial b^{0k}}{\partial w^{i}} + \frac{\partial\Gamma} 
{\partial b^{km}} \frac{\partial b^{km}}{\partial w^{i}} \nonumber \\
&=& \frac{1}{a} \frac{\partial \Gamma} {\partial b^{0i}}
\end{eqnarray}  
In the calculation of $\frac{\partial \delta \Gamma^{(1)}}{\partial 
b^{ij}}$ the following identities were used
\begin{equation}
\left. {\bar{\nabla}}^{k} {\beta}^{v(1)}_{k}\right|_{a=1} = -
\frac{1}{2} G^{km} {\beta}^{b^{\star}(1)}_{km}, \;\;\;\left.
{\bar{\nabla}}^{k} {\beta}^{w(1)}_{k}\right|_{a=1} = -
\frac{1}{2} F^{km} {\beta}^{b^{\star}(1)}_{km}.
\end{equation}

\section{\sc Variations of Anomaly Coefficients}
We now explain how to apply (3.12) in the context of this work. First for
ease of reference we quote from Ref.~\cite{kal}  all the relevant shifts in the
reduced fields
\begin{equation}
\delta a =  a a_{i} a^{i}+ \frac{1}{8} a^{2} F^{ij} F_{ij} 
+  \frac{1}{8} G^{ij} G_{ij},
\end{equation}
 \begin{equation}
\delta v_{i} = - \frac{1}{4} F_{i}^{\;\;k} a_{k} -
\frac{1}{8a} {{\hat{H}}}_{ikm} G^{km},
\end{equation}
\begin{equation}
\delta w_{i} = - \frac{1}{4} G_{i}^{\;\;k} a_{k} + \frac{1}{8} 
a {{\hat{H}}}_{ikm} F^{km},
\end{equation}
\begin{equation}
\delta {\hat{H}}_{ijk}= - \frac{3}{2} {\bar{\nabla}}_{[i}\left(
G_{j}^{\;\;m}
F_{k]m} \right) + 3 F_{[ij} \delta w_{k]} + 3 G_{[ij} \delta v_{k]}.
\end{equation}
\begin{equation}
\delta b_{ij} = \frac{1}{4} \left( G_{kj}F^{k}_{\;\;i} -
G_{ki}F^{k}_{\;\;j} \right) - v_{j} \delta w_{i} + v_{i} \delta w_{j}
\end{equation}
The variation for the metric $B$-function in tangent space is
\begin{eqnarray}
\delta B^{g}_{ab} &=& {e}_{a}^{\;\;\mu} {e}_{b}^{\;\;\nu} \delta B^{g}_
{\mu\nu} \nonumber \\
&=& {e}_{a}^{\;\;\mu} {e}_{b}^{\;\;\nu} \left[ \delta \lambda.
\frac{\partial}{\partial \lambda} {\beta}^{g}_{\mu\nu} - \mu \frac{d}{d
\mu} \delta g_{\mu\nu} \right]
\end{eqnarray}
The corrresponding equation for torsion follows directly. We now outline
the manipulation of these equations required to express them in terms of 
variations of $B$-functions for reduced fields.

The results for $B^{g}_{\hat{0}\hat{0}}$ and $B^{b}_{\hat{0}\alpha}$ are
achieved with ease since $e_{\hat{0}}^{\;\;i}=0$. We have
\begin{eqnarray}
\delta B^{g}_{\hat{0}\hat{0}} &=& e_{\hat{0}}^{\;\;0}e_{\hat{0}}^{\;\;0}
\delta B^{g}_{00} \nonumber \\
&=& \frac{1}{a}\delta B^{a} = \frac{1}{a} \left( \delta \lambda.
\frac{\partial}{\partial \lambda} {\beta}^{a} - \mu \frac{d}{d\mu}
\delta a \right), 
\end{eqnarray}
and
\begin{eqnarray}
\delta B^{b}_{\hat{0}\alpha} &=& e_{\hat{0}}^{\;\;0} {\bar{e}}_{\alpha}
^{\;\;i} \delta B^{w}_{i} \nonumber \\
&=& {\bar{e}}_{\alpha}^{\;\;i} \frac{1}{\sqrt{a}} \left( \delta
\lambda.\frac{\partial}{\partial \lambda} {\beta}^{w}_{i} - \mu
\frac{d}{d\mu}\delta w_{i} \right)
\end{eqnarray}

 The manipulations for $B^{g}_{\alpha\beta}$ and $B^{g}_{\hat{0}\alpha}$
are similar in style to each other. We now illustrate the simpler latter case.
It follows from (B6) that
\begin{equation} 
\delta B^{g}_{\hat{0}\alpha} = \frac{1}{\sqrt{a}}
{\bar{e}}_{\alpha}^{\;\;i} \left(\delta {\beta}^{g}_{oi} - v_{i}
\delta \beta^{a} \right).
\end{equation}
Since $g_{oi}=av_{i}$ we can compute 
\begin{equation}
\delta g_{oi} = v_{i} \delta a + a \delta v_{i}, 
\end{equation}  
and
\begin{equation}
{\beta}^{g}_{0i} = \beta^{a}v_{i} + a \beta^{v}_{i}.
\end{equation}
Hence we find
\begin{equation}
\delta {\beta}^{g}_{0i} = v_{i} \delta \beta^{a} + a \delta
\beta^{v}_{i},
\end{equation}
which upon substitution in (B.9) finally gives us
\begin{equation}
\delta B^{g}_{\hat{0}\alpha}= {\bar{e}}_{\alpha}^{\;\;i} \sqrt{a}
\delta B^{v}_{i}.
\end{equation}

Finally we illustrate the case for $B^{b}_{\alpha\beta}$. We need to
manipulate the expression for the change of the B-function  to calculate
$e_{\alpha}^{\;\;\mu} e_{\beta}^{\;\;\nu} X^{b^{\star}}_{\mu\nu}$ in 
(3.37). We have

\begin{equation}
\delta B^{b}_{\alpha\beta} = e_{\alpha}
^{\;\;\mu} e_{\beta}^{\;\;\nu} \left( \delta \lambda.\frac{\partial}
{\partial \lambda} \beta^{b}_{\mu\nu} - \mu \frac{d}{d\mu} \delta 
b_{\mu\nu} \right)
\end{equation}

We write out explicitly the first part of the right hand side of (B14)
\begin{equation}
e_{\alpha}
^{\;\;\mu} e_{\beta}^{\;\;\nu} \delta \lambda.\frac{\partial} {\partial
\lambda} \beta^{b}_{\mu\nu} = {\bar{e}}_{\alpha}^{\;\;i}{\bar{e}}_
{\beta}^{\;\;j} \left( \delta \lambda.\frac{\partial} {\partial
\lambda} \beta^{b}_{ij} - v_{i}\delta \lambda.\frac{\partial} {\partial
\lambda} \beta^{w}_{j} + v_{j} \delta \lambda.\frac{\partial} {\partial
\lambda} \beta^{w}_{i} \right). 
\end{equation}
Given the relation $B^{b^{\star} }_{ij} = B^{b}_{ij} -{v}_{i}B^{w}_{j} +
{v}_{j}B^{w}_{i}$ that we defined in (2.28), we can write (B15) in the
following form

\begin{equation}
e_{\alpha}
^{\;\;\mu} e_{\beta}^{\;\;\nu} \delta \lambda.\frac{\partial} {\partial
\lambda} \beta^{b}_{\mu\nu} = {\bar{e}}_{\alpha}^{\;\;i}{\bar{e}}_
{\beta}^{\;\;j} \left( \delta \lambda.\frac{\partial} {\partial
\lambda} \beta^{b^{\star}}_{ij} + \beta^{w}_{j} \delta v_{i} -
\beta^{w}_{i} \delta v_{j} \right)
\end{equation}  
where $\beta^{b^{\star}(1)}_{ij}$ may be read off from (2.33). 
The approach for calculating the remaining part of 
(B14) is straightforward and hence one can compute $e_{\alpha}
^{\;\;\mu} e_{\beta}^{\;\;\nu} X^{b^{\star}}_{\mu\nu}$.

\end{document}